\documentclass[english]{article}
\usepackage[latin9]{inputenc}
\usepackage{geometry}
\usepackage{verbatim}
\usepackage{amsbsy}
\usepackage{amssymb}
\usepackage{graphicx}

\makeatletter

\providecommand{\tabularnewline}{\\}

\usepackage{times}
\usepackage{epstopdf}

\date{}

\makeatother

\usepackage{babel}
\begin{document}

\title{Learning Rank Functionals: An Empirical Study}

\author{Truyen Tran, Dinh Phung, Svetha Venkatesh\\
Centre for Pattern Recognition and Data Analytics, Deakin University, Australia}
\maketitle
\begin{abstract}
Ranking is a key aspect of many applications, such as information
retrieval, question answering, ad placement and recommender systems.
Learning to rank has the goal of estimating a ranking model automatically
from training data. In practical settings, the task often reduces
to estimating a rank functional of an object with respect to a query.
In this paper, we investigate key issues in designing an effective
learning to rank algorithm. These include data representation, the
choice of rank functionals, the design of the loss function so that
it is correlated with the rank metrics used in evaluation. For the
loss function, we study three techniques: approximating the rank metric
by a smooth function, decomposition of the loss into a weighted sum
of element-wise losses and into a weighted sum of pairwise losses.
We then present derivations of piecewise losses using the theory of
high-order Markov chains and Markov random fields. In experiments,
we evaluate these design aspects on two tasks: answer ranking in a
Social Question Answering site, and Web Information Retrieval.
\end{abstract}
\global\long\def\x{\mathit{\mathbf{x}}}
\global\long\def\y{\mathit{\mathbf{y}}}
\global\long\def\z{\mathit{\mathbf{z}}}
\global\long\def\o{\mathit{\mathbf{o}}}
\global\long\def\w{\mathit{\mathbf{w}}}
\global\long\def\r{\mathit{\mathbf{r}}}
\global\long\def\perm{\mathit{\boldsymbol{\pi}}}
\global\long\def\param{\mathit{\boldsymbol{\theta}}}
\global\long\def\Real{\mathit{\mathbb{R}}}
\global\long\def\Risk{\mathit{\mathcal{R}}}
\global\long\def\wPL{\mbox{WPL}}
\global\long\def\rPL{\mbox{RPL}}
\global\long\def\wPLL{\mbox{WpLL}}
\global\long\def\wUB{\mbox{WUB}}

\section{Introduction}

Ranking is central to many applications in information retrieval,
question answering, online ad placement and recommender systems. Typically,
given a query (e.g. a set of keywords, a question, or an user), we
need to return a ranked list of relevant objects (e.g. a set of documents,
potential answers, or shopping items). Learning to rank (LTR) (e.g.
see \cite{liu2009learning} for a recent survey) is a machine learning
approach to automatically estimate a rank model from training data,
and offers promise way to leverage a wide prior knowledge. This includes
relevance to content, context or profile and object qualities such
as well-structuredness and authority. 

Most LTR algorithms aim at estimating a \emph{rank functional} $f(q,o)$
which takes a query $q$ and an object $o$ and returns a real score,
which is then used to rank objects with respect to the query. Typically
object order reflects its relevance of the object to the query, that
is, top ranked objects are considered most relevant. Mathematically,
this is the function estimation problem, but in the new setting of
ranking - where the goal is to output a sorted list of objects rather
than to compute the function itself. This is also different from the
traditional settings of regression or classification, where continuous
or discrete labels are predicted. 

This paper provides an empirical study to probe the construction of
rank functionals, and studies the contribution of the factors relevant
to the problem. Consider, for example, two application domains: answer
ranking in a Social Question Answering (SQA) site, and Web Information
Retrieval (WIR). In SQA, community members respond to the question
asked by another member, but the quality of answers varies greatly,
making answer ranking critical. Often, we have access to the actual
questions and answers, and in fact, questions can be quite rich and
contain deep linguistic structures. On the other hand, in Web document
retrieval, queries are often short, but there are many clues to assess
the quality of the related documents, for example, the link structures,
the authority of the domain and the organisation of the page. It is
safe to assume that there are probably hundreds of relevance indicators
currently employed in current major Web search engines. 

These differences suggest that the input representation for the rank
functionals can vary significantly from domain to domain. When the
query and objects are of different modalities, textual query versus
image objects for example, it is reasonable to represent the query
and objects separately. Likewise, in the case of SQA separate representation
can be applied to questions and answers since they are generally rich
in content and structure. On the other hand, for Web document retrieval,
prior knowledge of relevance indicators has been well-studied for
decades, and it can be useful to represent the query-object pair as
a combined vector.

The second aspect is the choice of the rank functionals, which captures
the relevancy of an object given a query. The specification of rank
functionals must be based on on the input representation, but it is
also critical that the functional space is rich enough to capture
data variability. For the separate query-object representations, we
study transformation methods to first project the query vector and
the object vector onto the same subspace and then combine them. For
the combined representation, we investigate the usefulness of quadratic
functions together with overfitting controls.

Third, learning often minimises some loss function based on training
data. It is therefore important that the loss functions reflect the
rank metrics that will be used to evaluate the algorithm at testing.
In particular, this involves placing more weight for high relevance
scores and the first few objects at the top of the rank list. The
main issue is that rank metrics are often computed based on the all
objects of the query, and they are discrete in nature, making it difficult
to optimise directly. To this end, we study different choices for
making a loss function closely related to the rank errors while maintaining
smoothness. In particular, we examine methods to approximate rank
metrics, and investigate weighting schemes to combine piecewise rank
losses, where the purpose of the weights are to emphasize the high
rating and discount for the low rank. 

Next, we study probabilistic approaches to derive piecewise rank losses.
The first approach is based on the theory of high-order Markov chains.
In particular, we investigate the utility of a weighted version of
the Plackett-Luce model \cite{luce1959individual}\cite{plackett1975analysis}
and its reverse. The second relies on Markov random fields (MRFs),
where we suggest the weighted version of the pseudo-likelihood, as
well as propose a piecewise approximation to the intractable MRFs.
We prove that this piecewise approximation provides an upper-bound
on the log-loss of the original MRF.

Finally, we evaluate these design aspects in the domain of Social
Question Answering (SQA) with an autism dataset retrieved from the
Yahoo! QA site, and the domain of Web Information Retrieval (WIR)
with the dataset from the Yahoo! LTR challenge \cite{chapelle2011yahoo}.
In SQA, we apply the separate representation of features - one for
questions and one for answers - as input for rank functionals. We
investigate two nonlinear rank functionals against three losses: the
multiclass logistic, the loss based on approximating the MRR metric,
and the pairwise logistic loss. The results shown that it is better
to use the loss specifically designed for the - the multiclass logistic
loss in this case.

Differing from the SQA, the WIR data has pre-computed features in
the combined representation. The hypothesis is that we can use the
quadratic rank functionals to discover predictive feature conjunctions.
We find that when regularisation and overfitting controls are properly
installed, second-order features are indeed more predictive than linear
counterparts. Another difference from the SQA data is that, the WIR
data contain relevance ratings in a small numerical scale, leading
to frequent occurrence of ties. This suggests the use of group-level
rank functionals, where objects of the same rating are grouped into
a mega-object. We evaluate several aggregation methods for computing
the group-level rank functionals. The results indicate that max and
geometric mean aggregations can be competitive, while maintaining
modest computational requirements due to the small number of resultant
groups. Experiments on the WIR data also show consistent results -weighting
is critical for many piecewise loss functions. In particular, for
the element-wise decomposition of rank losses, rank discount weighting
is the most influential, while for the pairwise decomposition, rating
difference combined with query length normalisation is the most effective
method. 

This paper is organised as follows. Section~2 presents a holistic
picture of the problem of LTR, and the specification details of rank
functionals as well as loss functions. In particular, we discuss the
piecewise weighting schemes to approximate the loss functions. Section~3
follows by describing probabilistic approaches to derive piecewise
losses. Design issues identified in the paper are then evaluated in
Section~4, where we present the experiments on the Yahoo! QA data,
and the Yahoo! LTR challenge data. Section~5 provides further discussion
on these aspects, followed by related work in Section~6. Finally
Section~7 concludes the paper.

\section{Estimating Rank Functionals }

In a typical setting, given a query $q$ and a set of related objects
$\o^{(q)}=(o_{1},o_{2},..,o_{N_{q}})$ we want to output a rank list
of objects. Ideally, this would means estimating an optimal permutation
$\perm^{(q)}=(\pi_{1},\pi_{2},..,\pi_{N_{q}})$ so that $\pi_{i}<\pi_{j}$
whenever $o_{i}$ is preferred to $o_{j}$, i.e.$o_{i}\succ o_{j}$.
However, this is often impractical since the number of all possible
permutations is $N_{q}!$. A more sensible strategy would be estimating
a real \emph{rank functional} $f(q,o_{i};\o_{\neg i}^{(q)})$ so that
$f(q,o_{i};\o_{\neg i}^{(q)})>f(q,o_{j};\o_{\neg j}^{(q)})$ whenever
$o_{i}\succ o_{j}$, where $\o_{\neg i}^{(q)}$ denotes all objects
for query $q$ except for $o_{i}$. This is efficient since the cost
of a typical sorting algorithm is $N_{q}\log N_{q}$.

For simplicity, in this paper we drop the explicit dependency between
$o_{i}$ and other objects, and write the rank functional as $f(q,o_{i})$.
In training data, we are given for each query a labelling scheme $\r^{(q)}$.
This labelling can be a rank list $\perm^{(q)}$, but more often,
it is a set of relevance scores, i.e. $\r^{(q)}=(r_{1},r_{2},...,r_{N_{q}})$
where $r_{i}$ is typically a small integer. Given a training data
of $D$ queries, the general way of estimating $f(q,o_{i})$ is to
minimise an regularised empirical risk functional
\begin{equation}
\Risk(f)=\frac{1}{D}\sum_{q=1}^{D}\ell(\r^{(q)},\{f(q,o_{i})\}_{i=1}^{N_{q}})+\lambda\Omega(f)\label{eq:risk}
\end{equation}
with respect to $f$, where $\ell(\r^{(q)},\{f(q,o_{i})\}_{i=1}^{N_{q}})$
is the loss function, $\Omega(f)$ is a (convex) regularisation function
and $\lambda>0$ is the regularisation factor.

\begin{table}
\begin{centering}
\begin{tabular}{lc}
\hline 
\noalign{\vskip\doublerulesep}
MRR  & $\frac{1}{D_{test}}\sum_{q=1}^{D_{test}}\frac{1}{\hat{\pi}{}_{*}}$\tabularnewline[\doublerulesep]
\noalign{\vskip\doublerulesep}
NDCG@T & $\frac{N@T}{N_{max}@T}$ where $N@T=\sum_{i=1}^{T}\frac{2^{r_{i}}-1}{\log_{2}(1+\hat{\pi}{}_{i})}$ \tabularnewline[\doublerulesep]
\noalign{\vskip\doublerulesep}
ERR & $\sum_{i}\frac{1}{\hat{\pi}_{i}}R(r_{i})\prod_{j|\hat{\pi}_{j}<\hat{\pi}_{i}}(1-R(r_{j}))$ \tabularnewline
 & ~~~~~~~~~~where $R(r_{i})=\frac{2^{r_{i}}-1}{16}$\tabularnewline
\hline 
\end{tabular}
\par\end{centering}

\caption{Evaluation metrics. Here $\hat{\pi}_{i}$ is the predicted position
of object $o_{i}$ and $\hat{\pi}{}_{*}$ is the predicted position
of the best object. \label{tab:Evaluation-metrics.}}
\end{table}

We now discuss details of rank functional $f(.)$ and rank loss $\ell(.)$.

\subsection{Specifying Rank Functionals $f$}

Depending on specific applications, for each query-object pair $(q,o_{i})$
we can represent the query and the object separately (e.g. as feature
vectors $\mathbf{y}_{q}\in\mathbb{R}^{n_{1}}$ and $\mathbf{z}_{o}\in\mathbb{R}^{n_{2}}$),
or combined into a single vector $\x_{i}^{(q)}\in\mathbb{R}^{m}$.
For example, in domains where queries and objects are of different
modalities, such as textual queries for image objects, separate representations
will be particularly useful. On the other hand, in situations where
rich domain knowledge can be utilised to obtain multiple relevance
and quality indicators, the the combined representation may be of
advantage.

\subsubsection{Separate Representation}

Since the two vectors $\y_{q}$ and $\z_{o}$ are from different spaces,
it can be useful to apply transformations into the same space. For
simplicity, let us focus on linear transformation operators $\mathbf{A}\in\mathbb{R}^{d\times n_{1}},\mathbf{B}\in\mathbb{R}^{d\times n_{2}}$
to convert $\{\y_{q},\z_{o}\}$ into $\{\mathbf{A}\mathbf{y}_{q},\mathbf{B}\mathbf{z}_{o}\}\in\Real^{d}$,
respectively. Given the transformations, we can compute the degree
of association using
\begin{equation}
f_{1}(\y_{q},\z_{o})=\boldsymbol{\sigma}(\mathbf{A}\mathbf{y}_{q})'\mathbf{B}\mathbf{z}_{o}\label{eq:neural-functionals}
\end{equation}
where $\boldsymbol{\sigma}$ is an element-wise mapping. Typically
we choose the sigmoid or tanh functions to introduce nonlinearity.

Another option is to use the distance metric
\begin{equation}
f_{2}(\y_{q},\z_{o})=-\frac{1}{\tau}\left\Vert \mathbf{A}\mathbf{y}_{q}-\mathbf{B}\mathbf{z}_{o}\right\Vert ^{2}\label{eq:metric-functionals}
\end{equation}
 for some scaling factor $\tau>0$. Thus, learning the rank functional
$f$ reduces to estimating the transformation matrices $\mathbf{A}$
and $\mathbf{B}$.

\subsubsection{Combined Representation}

We will focus on quadratic functions for the case of combined representation.
A quadratic rank model is a function of pairwise feature conjunction/interaction%
{} of the form 
\begin{equation}
f_{3}(q,o)=\alpha_{0}+\w'\x+\x'\mathbf{C}\x\label{eq:quadratic-functionals}
\end{equation}
where $\x\in\Real^{m}$ is the feature vector, and $\w\in\Real^{m}$,
$\mathbf{C}\in\Real^{m\times m}$ are parameters. The hypothesis is
that certain feature conjunction/interactions are likely to emphasize
some relevancy/quality aspects captured in the feature engineering
process. To ensure that such second-order features will benefit rather
than harm the performance, we need to remove unlikely combinations.
One way is to impose sparsity-boosting regularisation function $\Omega(f)$
but it is likely to complicate the optimisation process. In this study,
we follow a simpler approach by pre-filtering unlikely second-order
features by some criteria. In particular, we apply Pearson's correlation%
\footnote{For two variables $u,v$, the Pearson's correlation is computed as
$c(u,v)=\frac{\sum_{i}(u_{i}-\bar{u})(v_{i}-\bar{v})}{\sqrt{\sum_{i}(u_{i}-\bar{u})^{2}}\sqrt{\sum_{i}(v_{i}-\bar{v})^{2}}}$,
where $\bar{u},\bar{v}$ are the mean values of $u$ and $v$, respectively.%
} between second-order features and relevance scores as a rough measure
of quality. Then we filter out second-order features whose absolute
correlation coefficient is less than a certain threshold $\rho\in(0,1)$.

\begin{table}
\begin{centering}
\begin{tabular}{|l|c|}
\hline 
 & $\bar{f}(\{o_{j}|r_{j}=\bar{r}\})$\tabularnewline
\hline 
Min & $\min\{f(o_{j})|r_{j}=\bar{r}\}$\tabularnewline
Max & $\max\{f(o_{j})|r_{j}=\bar{r}\}$\tabularnewline
Arithmetic Mean & $\frac{1}{K}\sum_{j}f(o_{j})$\tabularnewline
Geometric Mean & $\log\left(\frac{1}{K}\sum_{j}\exp f(o_{j})\right)$\tabularnewline
\hline 
\end{tabular}
\par\end{centering}

\caption{Group-level rank functions.\label{tab:Group-level-rank-functions.}}
\end{table}

\subsubsection{Group-level Functions}

In specifying $f$, one issue which is often overlooked is the the
presence of ties. In practice, often the data is given in the form
of relevance ratings which results in many ties. Thus, it may be useful
to operate directly at the rating level, where ties are incorporated.
The idea here is to treat a group of objects with the same rating
as a mega-object. Like individual objects, a mega-object has a group-level
rank function defined upon. This group-level function is an aggregation
function, which combines individual rank functions in a sensible manner.
For example, it is necessary to normalise the group-level function
since the group size can vary greatly. Denote by $\bar{f}(\{o_{j}|r_{j}=\bar{r}\})$
the group-level rank function for the group whose relevance label
is $\bar{r}$. Table~\ref{tab:Group-level-rank-functions.} lists
several candidate aggregation functions. Thus, assuming the relevance
ratings are discrete with $|L|$ levels, there are at most $|L|$
groups. We note in passing that group-level function can be in conjunction
with any individual functions.

Group-level rank functions can be useful for several reasons. First,
they implicitly impose prior knowledge of ties. Second, for functions
that perform averaging, we may achieve some kind of regularisation,
making it more robust to noise. Third, often group-level functions
can be computed in linear time and the overall complexity is also
linear in number of objects per query. Therefore, we have computational
savings, e.g. against pairwise losses, whose complexity is often quadratic
in query size.

\subsection{Specifying Rank Loss $\ell$}

The next important issue is the specification of the loss function
$\ell$. For clarity, let us make use of the notation $\ell(\r,\o)$
to refer to the loss function and we drop the explicit mention of
the query $q$ when no confusion occurs. It is reasonable to assume
that $\ell(\r,\o)$ is highly correlated with the rank metrics used
in evaluation, that is, a low loss should lead to high performance
according to the rank metrics. Typically, rank metrics used in practice,
such as MRR, NDCG \cite{jarvelin2002cumulated} and ERR \cite{chapelle2009expected}
(see Table~\ref{tab:Evaluation-metrics.}), are functions of predicted
positions of objects in the rank list. The predicted position is estimated
as $\hat{\pi}_{i}=1+\sum_{j\ne i}\delta[f(o_{j})>f(o_{i})]$. This
suggests that it may be useful for the loss function to incorporate
pairwise comparisons among objects.

In this paper, we focus on two situations: when we need to emphasize
on the singe best object, and when several objects are relatively
useful.

\subsubsection{Multiclass Logistic Loss}

The first situation reduces to the multiclass categorisation problem,
where a popular choice is the multiclass logistic loss
\begin{equation}
\ell_{1}(\r,\o)=-\log P(r^{*}=1|\o)\label{eq:multi-logit}
\end{equation}
where $P(r_{i}=1|\o)=\exp\{f(q,o_{i})\}/\sum_{j=1}^{N}\exp\{f(q,o_{j})\}$,
and $P(r^{*}=1|\o)$ denotes the probability of that best object is
ranked first. 

The second situation is more general, but also requires more sophisticated
techniques to design the loss $\ell(\r,\o)$. In this paper, we focus
on three techniques: approximating the rank error, decomposition of
loss into element-wise weighted sum and into pairwise weighted sum.

\subsubsection{Smoothing Rank Errors}

The first technique is to approximate the indicator $\delta[f(o_{i})>f(o_{j})]$
by a smooth function $\varrho(f(o_{i}),f(o_{j}))\in[0,1]$, and then
optimise the resulting metrics directly
\begin{equation}
\ell_{2}(\r,\o)=1-M\left(\r,\{\varrho(f(o_{i}),f(o_{j})\}_{ij|j>i}\right)\label{eq:approx-metric-loss}
\end{equation}
where $M(.)$ is rank metric. One popular choice of $\varrho(.)$
is the sigmoid function
\begin{equation}
\varrho(f(o_{i}),f(o_{j}))=\frac{1}{1+\exp\{-(f(o_{i})-f(o_{j}))\}}.\label{eq:sigmoid-step-func}
\end{equation}
The result is a smooth loss function with respect to $f$. The drawback
is that the loss function is not flexible though complex. When the
sigmoid function approaches the step function, it is hard to improve
the performance any further.

\subsubsection{Element-wise Decomposition}

The second technique is to approximate the query-level loss by the
following element-wise sum
\begin{equation}
\ell_{3}(\r,\o)=\sum_{i=1}^{N}W_{i}\omega(r_{i},o_{i};\r_{\neg i},\o_{\neg i})\label{eq:pointwise}
\end{equation}
where $\omega(r_{i},o_{i};\r_{\neg i},\o_{\neg i})$ is a function
of one variable $r_{i}$ while fixing other labellings $\r_{\neg i}$,
and $W_{i}\ge0$ is a weighting parameter possibly depending on true
position $\pi_{i}$ of the object $o_{i}$. An example is perhaps
in pointwise regression, where $\omega(r_{i},o_{i};\r_{\neg i},\o_{\neg i})=(r_{i}-f(o_{i}))^{2}$.
Interestingly, it has been proven that with appropriate weighting
scheme $W_{i}$, this approximation is indeed a good bound on the
NDCG metric \cite{cossock2008statistical}. Another example is $\omega(r_{i},o_{i}|\r_{\neg i},\o_{\neg i})=-\log P(r_{i}|\r_{\neg i},\o)$
as in the case of pseudo-likelihood approximation to the Markov random
field, as we will present in the next section.

\begin{table}
\begin{centering}
\begin{tabular}{|l|c|}
\hline 
 & $\varphi(r_{i},r_{j},o_{i},o_{j})$ for $o_{i}\succ o_{j}$\tabularnewline
\hline 
Quadratic  & $\{1-(f(q,o_{i})-f(q,o_{j}))\}^{2}$\tabularnewline
Hinge \cite{joachims2002optimizing} & $\max\{0,1-(f(q,o_{i})-f(q,o_{j}))\}$\tabularnewline
Exponential \cite{freund2004eba} & $\exp\{-(f(q,o_{i})-f(q,o_{j})\}$\tabularnewline
Logistic \cite{burges2005learning} & $\log(1+\exp\{-(f(q,o_{i})-f(q,o_{j})\})$\tabularnewline
\hline 
\end{tabular}
\par\end{centering}

\caption{Some pairwise losses.\label{tab:Pairwise-losses}}
\end{table}

\subsubsection{Pairwise Decomposition}

The third technique is combining \emph{piecewise loss}es, where each
piece is a bivariate function 

\begin{eqnarray}
\ell_{4}(\r,\o) & = & \sum_{i=1}^{N}\sum_{j>i}V_{ij}\varphi(r_{i},r_{j},o_{i},o_{j})\label{eq:pairwise}
\end{eqnarray}
where $V_{ij}\ge0$ is a weighting factor for the pair $\{i,j\}$,
possibly depending on the true positions $\{\pi_{i},\pi_{j}\}$. This
approximation is indeed well-studied in the LTR literature, where
$\varphi(r_{i},r_{j},o_{i},o_{j})$ is often the pairwise loss (see
Table~\ref{tab:Pairwise-losses} for popular losses). In fact, it
has been shown that unweighted hinge, exponential and logistic losses
are actually upper bounds of 1-NDCG \cite{chen22ranking}.

This paper studies a number of important issues regarding the design
of an effective LTR algorithm. First is the data representation of
the pair query-object $(q,o)$. Second is the choice of rank functionals
$f(q,o)$ which capture the importance and relevancy of object $o$
with respect to the query $q$. The effectiveness of the quadratic
functional is studied in the experimental section. Third, the design
of appropriate loss functions $\omega(.)$ and $\varphi(.)$ and specification
the weights $\{W_{i}\}$ and $\{V_{ij}\}$ to combine those loss functions.
It is likely that the weights should emphasize the high rating and
discount for the low rank.

\section{Deriving $\omega(.)$ and $\varphi(.)$ from Probabilistic Query
Models}

In this section, we present probabilistic approaches to derive the
functions $\omega(.)$ of Eq.(\ref{eq:pointwise}) and $\varphi(.)$
of Eq.(\ref{eq:pairwise}). We start from specifying the query-level
model distribution $P(\r|\o)$. We consider two model representations:
the \emph{high-order Markov chain} and the \emph{Markov random field}. 

\begin{figure*}
\begin{centering}
\begin{tabular}{cccc}
\includegraphics[width=0.2\textwidth]{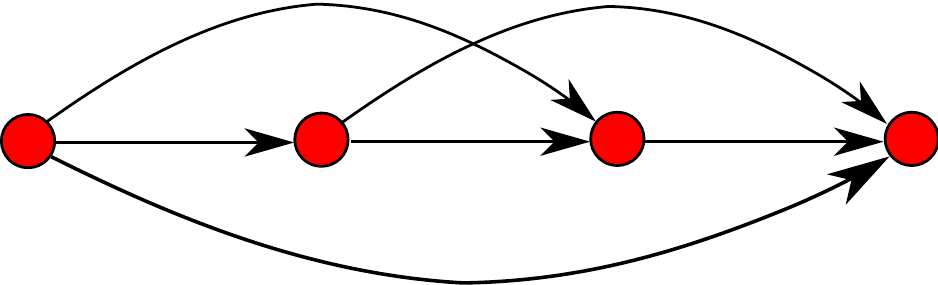}~ & \includegraphics[width=0.2\textwidth]{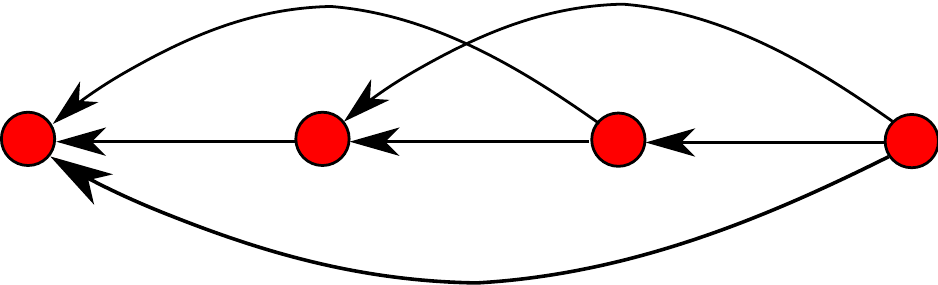}~ & \includegraphics[width=0.2\textwidth]{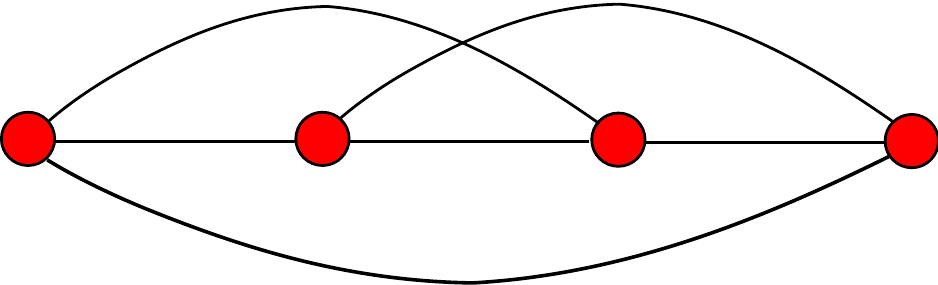}~ & \includegraphics[width=0.2\textwidth]{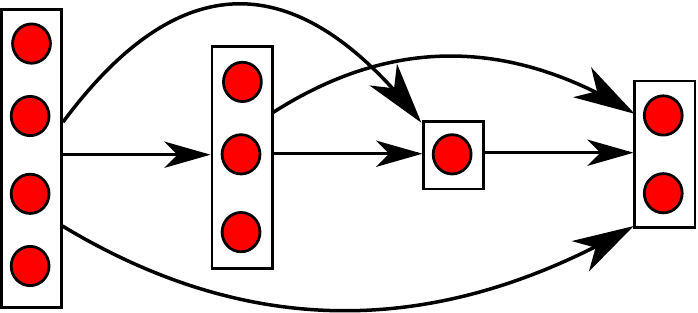}\tabularnewline
(a) & (b) & (c) & (d)\tabularnewline
\end{tabular}
\par\end{centering}

\caption{Graphical illustration of query models. (a) Plackett-Luce, (b) Reverse
Plackett-Luce, (c) Markov random field and (d) Group-level Plackett-Luce.\label{fig:Graphical-illustration} }
\end{figure*}

\subsection{High-order Markov chains}

For this model representation we work directly the permutation model
of objects. Thus we will use the permutation notation $\perm=(\pi_{1},\pi_{2},...,\pi_{N})$
to denote the labelling scheme, where $\pi_{i}$ is the position of
the object in the list according to the permutation $\perm$. Hence,
according to $\perm$, $\pi_{i}<\pi_{j}$ means that the object $o_{i}$
is ranked higher than the object $o_{j}$, or equivalently $o_{i}\succ o_{j}$.
In essence, the model aims at specifying the permutation distribution
$P(\perm|\o)$.

For clarity, we drop the explicit reference to $\o$ and write $P(\perm)$
as well as $\ell(\perm)$ instead of $P(\perm|\o)$ and $\ell(\perm,\o)$,
respectively. Let us start from the probabilistic theory that any
joint distribution of $N$ variables can be factorised according to
the chain-rule as follows

\begin{equation}
P(\pi_{1},\pi_{2},...,\pi_{N})=P(\pi_{1})\prod_{i=2}^{N}P(\pi_{i}|\boldsymbol{\pi}_{1:i-1})\label{eq:PL-factorisation}
\end{equation}
where $\boldsymbol{\pi}_{1:i-1}$ is a shorthand for $(\pi_{1},\pi_{2},..,\pi_{i-1})$.
See Figure~\ref{fig:Graphical-illustration}a for a graphical representation
of this factorisation. 

Informally in the context of LTR, this factorisation can be interpreted
as follows: choose the first object in the list with probability of
$P(\pi_{1})$, and choose the second object with probability of $P(\pi_{2}|\perm_{1})$,
and so on. The Luce's axioms of choice \cite{luce1959individual}
assert that we should choose an object with probability proportional
to its \emph{worth}. Translated into our LTR problem, the worth of
the object $o_{i}$ can be defined%
\footnote{Worth can be any positive function as long as it is monotonically
increasing in $f$%
} as $\phi(o_{i})=\exp\{f(o_{i})\}$. Any finally, according to Plackett
\cite{plackett1975analysis}, we can define the conditional probabilities
as follows 
\begin{eqnarray*}
P(\pi_{1}) & = & \frac{\phi(o_{1})}{\sum_{j=1}^{N}\phi(o_{j})}\\
P(\pi_{i}|\boldsymbol{\pi}_{1:i-1}) & = & \frac{\phi(o_{i})}{\sum_{j=i}^{N}\phi(o_{j})}
\end{eqnarray*}
This model is called the Plackett-Luce (PL) and has been studied in
the context of LTR in \cite{xia2008listwise_FULL} .

\subsubsection{Weighted PL ($\wPL$)}

Typically, we learn the PL model by maximising the data likelihood,
or equivalently, minimising the negative log-likelihood $-\log P(\perm)$.
However, this does not necessarily lead to good performance since
the likelihood and a typical rank metric are different. For example,
rank metrics often put a great emphasis on the first few objects and
effectively ignore the rest. This suggests a weighted log-loss
\begin{equation}
\ell_{5}(\perm)=-W_{1}\log P(\pi_{1})-\sum_{i=2}^{N}W_{i}\log P(\pi_{i}|\boldsymbol{\pi}_{1:i-1})\label{eq:weighted-PL}
\end{equation}
which is in the form of Eq.(\ref{eq:pointwise}).

\subsubsection{Reverse PL ($\rPL$)}

First, note that the factorisation in Eq~\ref{eq:PL-factorisation}
is general, and is not unique. In fact if we permute the indices of
objects, the factorisation can still hold. Second, we are free to
choose any realisation of the conditional distribution $P(\pi_{i}|\boldsymbol{\pi}_{1:i-1})$.
For example, we can derive a \emph{reverse} Plackett-Luce model as
follows (see Figure~\ref{fig:Graphical-illustration}b for a graphical
representation) 
\begin{equation}
Q(\pi_{1},\pi_{2},...,\pi_{N})=Q(\pi_{N})\prod_{i=1}^{N-1}Q(\pi_{i}|\boldsymbol{\pi}_{i+1:N})\label{eq:PL-reverse}
\end{equation}

The tricky part is to define $Q(\pi_{N})$ and $Q(\pi_{i}|\boldsymbol{\pi}_{i+1:N})$.
Since $\pi_{N}$ is interpreted as the most irrelevant object in the
list, $Q(\pi_{N})$ cannot be thought as the probability of choosing
object according to its worth. We think of $Q(\pi_{N})$ as the probability
of \emph{eliminating} the object instead. Thus, it is reasonable to
assume that \emph{the probability of an object being eliminated is
inversely proportional to its worth.} This suggests the following
specifications

\begin{eqnarray*}
Q(\pi_{N}) & = & \frac{\phi^{-1}(o_{N})}{\sum_{j=1}^{N}\phi^{-1}(o_{j})}\\
Q(\pi_{i}|\boldsymbol{\pi}_{i+1:N}) & = & \frac{\phi^{-1}(o_{i})}{\sum_{j=1}^{i}\phi^{-1}(o_{j})}.
\end{eqnarray*}
This factorisation in fact has an interesting interpretation. First,
we choose the last object to eliminated, then choose the second one,
and so on. Note that, due to specific choices of conditional distributions,
$P\ne Q$ in general. Again, position-based weighting can be applied
\begin{equation}
\ell_{6}(\perm)=-W_{N}\log Q(\pi_{N})-\sum_{i=1}^{N-1}W_{i}\log Q(\pi_{i}|\boldsymbol{\pi}_{i+1:N}),\label{eq:weighted-reverse-PL}
\end{equation}
i.e. this is an instance of Eq.(\ref{eq:pointwise}).

\subsection{Markov random fields}

In this model representation, we consider the case where discrete
ratings $\r=(r_{1},r_{2},...,r_{N})$ are given to corresponding objects
$\o=(o_{1},o_{2},..,o_{N})$, where each $r_{i}$ is drawn from the
same label set $L$. We treat the ratings as random variables and
aim at modelling a joint distribution $P(\r|\o)$. Again, for clarity,
we will drop the explicit reference to $\o$ and write $P(\r)$ as
well as $\ell(\r)$ instead of $P(\r|\o)$ and $\ell(\r,\o)$, respectively. 

In particular, the model has the following form

\begin{eqnarray}
P(\r) & = & \frac{1}{Z}\prod_{i}\prod_{j>i}\psi(r_{i},r_{j})\,\,\mbox{where}\label{eq:MRF}\\
\psi(r_{i},r_{j}) & = & \exp\{\gamma\sum_{i}\sum_{j>i}\mbox{sign}[r_{i}-r_{j}](f(o_{i})-f(o_{j}))\}\nonumber \\
Z & = & \sum_{r_{1},r_{2}...,r_{N}}\prod_{i}\prod_{j>i}\psi(r_{i},r_{j})
\end{eqnarray}
$\gamma>0$ is the scaling factor accounting for the variation in
number of objects per query (e.g. $\gamma=\frac{2}{N(N-1)}$). The
idea is that the model encourages the correlation between the relative
rating order (encoded in $\mbox{sign}[r_{i}-r_{j}]$) and the relative
ranking order (by $f(o_{i})-f(o_{j})$). In fact this idea is very
similar to that of pairwise models, except we now consider all pairs
simultaneously. The result is a fully connected Markov random field%
\footnote{To be more precise, the model should be called conditional MRF, or
conditional random fields since state variables are conditioned on
the objects.%
} (MRF, see Figure~\ref{fig:Graphical-illustration}c for a graphical
representation). 

However, it is well-known that learning in MRFs is difficult since
the likelihood can not be computed exactly in general cases. In practice,
we often resort to approximate methods, such as MCMC sampling. Here
we propose several approximation alternatives, where the smoothness
of the approximated loss is retained, making optimisation easier.

\subsubsection{Weighted Pseudo-Likelihood ($\wPLL$)}

The first approximation is to work on the local conditional distribution
$P(r_{i}|\r_{\neg i})$ instead of $P(\r)$ directly, where 
\[
P(r_{i}|\r_{\neg i})=\frac{\prod_{j\ne i}\psi(r_{i},r_{j})}{\sum_{r_{i}'}\prod_{j\ne i}\psi(r_{i}',r_{j})}.
\]
We propose the following weighted pseudo-likelihood

\begin{eqnarray}
\ell_{7}(\r) & = & -\sum_{i}W_{i}\log P(r_{i}|\r_{\neg i})\label{eq:weight-pseudo}
\end{eqnarray}
which has the form of Eq.(\ref{eq:pointwise}).

\subsubsection{Weighted Pairwise Upper-Bound ($\wUB$)}

We start by observing that \emph{the log-loss $\ell(\r)=-\log P(\r)$
can be upper-bounded by a sum of pairwise losses
\begin{equation}
\ell(\r)\le-\sum_{i}\sum_{j>i}\log Q(r_{i},r_{j})+C\label{eq:upper-bound}
\end{equation}
for some constant $C$ and $Q(r_{i},r_{j})\propto\psi(r_{i},r_{j})$.} 

The proof is presented in the appendix, but interested readers may
start from the general H\"{o}lder's inequality and noting that the
normalising constant in Eq.(\ref{eq:MRF}) $Z=\sum_{\r}\prod_{i}\prod_{j>i}\psi(r_{i},r_{j})$
has the sum-product form. Again, to account for the rank metrics,
we propose to use the following weighted loss
\begin{equation}
\ell_{8}(\r)=-\sum_{i}\sum_{j>i}W_{ij}\log Q(r_{i},r_{j})\label{eq:weighted-piecewise}
\end{equation}
which has the form of Eq.(\ref{eq:pairwise}).

\section{Experimental Studies}

In this section, we evaluate the design choices on two applications:
\emph{Social Question Answering }and\emph{ Web Information Retrieval}.
The first case involves ranking answers offered to a question in a
Social QA site whilst the second case aims at ranking documents related
to a query.

Before proceeding into the experimental details, let us make a note
on the implementation. Until now, we have not presented the details
of the learning process where parameters are estimated by minimising
the empirical risk $\Risk(f)$ in Eq.\ref{eq:risk}. All of the models
studied in this paper is continuous in the rank functional $f$, which
is continuous in model parameters. Thus the empirical risk is also
continuous in model parameters, where gradient-based optimisation
routines can be applied for find the minimisers of $\Risk(f)$. Since
the gradient details of mentioned models are not the main subject
of this paper, we omit for clarity. For all models, however, are regularised
by a simple Gaussian prior in parameters, and are trained using the
L-BFGS, a limited memory Newton-like algorithm.

\subsection{Answer Ranking in Social Question Answering}

QA offers a different perspective from traditional IR as the question
is often semantically rich, and the style in posing a question and
an answer is different. From (Yahoo! QA%
\footnote{http://answers.yahoo.com%
}) we collect all questions relevant to the topic of \emph{autism}.
In the training data, an answer is chosen as the best one for the
question. We use $36,113$ questions ($215,230$ answers) for training,
$1,000$ questions ($6,342$ answers) for development, and $999$
questions ($7,636$ answers) for testing.

\begin{table}
\begin{centering}
\begin{tabular}{c|c|c|c}
Rank func. & Multi.logit.$(\ell_{1})$ & MRR Opt.($\ell_{2}$) & Pair.logit.($\ell_{3}$)\tabularnewline
\hline 
$f_{1}$(Eq.\ref{eq:neural-functionals}) & 0.5474 & 0.4944 & 0.5360\tabularnewline
$f_{2}$(Eq.\ref{eq:metric-functionals}) & 0.5381 & 0.4662 & 0.4697\tabularnewline
\end{tabular}
\par\end{centering}

\caption{MRR scores of rank functionals for answer ranking, where $d=10$,
$n_{1}=n_{2}=3,000$. Multiclass logistic loss is defined in Eq.(\ref{eq:multi-logit}),
pairwise logistic loss is listed in Table~\ref{tab:Pairwise-losses},
and the MRR metric is in Table~\ref{tab:Evaluation-metrics.}. \label{tab:Rank-functionals-for-QA}}
\end{table}

We use words as features and employ the separate representation of
input and the rank functionals developed in Eqs.(\ref{eq:neural-functionals})
and (\ref{eq:metric-functionals}). We construct the vocabularies
for questions and answers separately. The vocabularies are constructed
by selecting the top $3,000$ frequent non-stop words in the training
questions, and training answers, respectively. 

Table~\ref{tab:Rank-functionals-for-QA} reports the MRR scores of
the two rank functionals when used in conjunction with three loss
functions: the multiclass logistic, the pairwise logistic and the
direct optimisation of the MRR metric. Note that since this data has
only one correct answer per query, it is essentially an instance of
the multiclass categorisation problem. In all combinations, the embedding
space has the dimensionality of $d=10$. For this problem, it appears
that the multiclass logistic method performs best, and optimising
the metric directly does not necessarily lead to improvement.

\subsection{Web Information Retrieval}

In this application, we employ data from the Yahoo! learning to rank
challenge \cite{chapelle2011yahoo}. The data contains the groundtruth
relevance scores (from $0$ for irrelevant to $4$ for perfectly relevant)
of $473,134$ documents returned from $19,944$ queries. We use a
subset of $1,568$ queries ($47,314$ documents) for training, and
another subset of $1,520$ queries ($47,313$ documents) for testing.
There are $519$ pre-computed unique features for each query-document
pair. Thus, this falls into the category of combined feature representation.
We first normalise the features across the whole training set to have
mean $0$ and standard deviation $1$ - this appears to consistently
improve the performance over raw features. When there is not enough
space, we report the ERR score only since it was required for the
challenge.

\subsubsection{Evaluation of Rank Functionals}

\textbf{Quadratic rank functionals.} We evaluate the quadratic rank
functionals $f_{3}$ (Eq.\ref{eq:quadratic-functionals}) with weighted
pairwise logistic loss (see Table~\ref{tab:Pairwise-losses}, $\ell_{4}$,
Eq.\ref{eq:pairwise}), where where the weight is $\left|2^{r_{i}-1}-2^{r_{j}-1}\right|/N_{q}$
for the pair $\{o_{i},o_{j}\}$. Table~\ref{tab:Performance-of-second-order}
reports the performance of this setting with respect to the feature
selection threshold $\rho$. It can be seen that quadratic rank models
can improve over linear models given appropriate overfitting control
(since the number of parameters is much larger than the linear case).
However, the improvement is not free - the time and space complexity
generally scale linearly with the number of parameters. Evaluation
with other losses shows the same pattern.\\

\begin{table}
\begin{centering}
\begin{tabular}{c|r|c|c|c|c}
$\rho$ & No. feats & N@1 & N@5 & N@10 & ERR\tabularnewline
\hline 
 & \emph{519} & \emph{0.6941} & \emph{0.6638} & \emph{0.6848} & \emph{0.4979}\tabularnewline
\hline 
0.00 & 134,477 & 0.6792 & 0.6480 & 0.6742 & 0.4889\tabularnewline
0.05 & 53,394 & 0.6979 & 0.6630 & 0.6862 & 0.4989\tabularnewline
0.10 & 26,318 & 0.7020 & 0.6651 & 0.6887 & 0.5019\tabularnewline
0.15 & 14,425 & 0.7090 & 0.6666 & 0.6897 & \textbf{0.5048}\tabularnewline
0.20 & 7,114 & \textbf{0.7112} & \textbf{0.6716} & \textbf{0.6931} & 0.5038\tabularnewline
0.25 & 2,983 & 0.6942 & 0.6636 & 0.6877 & 0.4991\tabularnewline
0.30 & 1,294 & 0.6926 & 0.6519 & 0.6730 & 0.4960\tabularnewline
\end{tabular}
\par\end{centering}

\caption{Performance of weighted pairwise logistic loss ($\ell_{4}$) with
second-order features w.r.t. the selection threshold $\rho$. For
comparison, the first row includes the result with first-order features.
\label{tab:Performance-of-second-order}}
\end{table}

\begin{table}
\begin{centering}
\begin{tabular}{l|c|c|c|c}
Aggregation type & N@1 & N@5 & N@10 & ERR\tabularnewline
\hline 
 & \emph{0.6796} & \emph{0.6518} & \emph{0.6777} & \emph{0.4845}\tabularnewline
\hline 
Min & 0.6027 & 0.6098 & 0.6430 & 0.4561\tabularnewline
Max & 0.6875 & 0.6555 & 0.6796 & 0.4961\tabularnewline
Arithmetic Mean & 0.6773 & 0.6465 & 0.6727 & 0.4892\tabularnewline
Geometric Mean & 0.6855 & 0.6539 & 0.6784 & 0.4918\tabularnewline
\end{tabular}
\par\end{centering}

\caption{Results of group-level rank functionals (see Table~\ref{tab:Group-level-rank-functions.}
for definitions), applied for unweighted Plackett-Luce loss ($\ell_{5}$
with all unity weights). For comparison, the top row includes the
result of the standard object-level rank functional. \label{tab:Results-of-group-level}}
\end{table}

\textbf{Group-level rank functionals.} Table~\ref{tab:Results-of-group-level}
reports the results of using group-level rank functionals in comparison
with standard linear ones, where the loss is the standard Plackett-Luce.
It can be seen that the group-level rank functionals (except for the
min-aggregation type) are competitive in the ERR scores.

\subsubsection{Evaluation of Rank Losses}

We now present the results for various rank losses, all trained on
linear rank functionals for simplicity.\\

\textbf{Metrics optimisation.} Table~\ref{tab:Approximate-metrics}
reports the results of approximating rank metrics of NDCG and ERR
using soft step functions in Eq.(\ref{eq:sigmoid-step-func}). The
MRR is not included since it assumes that there is only one best object
per query. It can be seen that the performance of rank metrics optimisation
is comparable to standard pairwise models (see Table~\ref{tab:Pairwise-losses}).\\

\begin{table}
\begin{centering}
\begin{tabular}{l|c|c|c|c}
 & N@1 & N@5 & N@10 & ERR\tabularnewline
\hline 
Pairwise Quadratic & 0.6608 & 0.6462 & 0.6712 & 0.4789\tabularnewline
Pairwise Logistic & 0.6617 & 0.6473 & 0.6717 & 0.4801\tabularnewline
Pairwise Hinge & 0.6573 & 0.6462 & 0.6718 & 0.4787\tabularnewline
\hline 
NDCG Optimisation & 0.6676 & 0.6439 & 0.6712 & 0.4871\tabularnewline
ERR Optimisation & 0.6683 & 0.6355 & 0.6606 & 0.4871\tabularnewline
\end{tabular}
\par\end{centering}

\caption{Approximate metrics optimisation using $\ell_{2}$ (see Table~\ref{tab:Evaluation-metrics.}
for metric definitions). N@$K$ is a short hand for NDCG score for
the top $K$ results.\label{tab:Approximate-metrics}}
\end{table}

\begin{table}
\begin{centering}
\begin{tabular}{c|c|c|c}
Weight $W_{i}^{(q)}$ & $\wPL$ & $\rPL$ & $\wPLL$\tabularnewline
\hline 
$1$ & \emph{0.4845} & \emph{0.4980} & \emph{0.4882}\tabularnewline
$r_{i}$ & 0.4955 & 0.4944 & 0.4980\tabularnewline
$\sqrt{r_{i}}$ & 0.4915 & 0.4957 & 0.4979\tabularnewline
$\frac{2^{r_{i}-1}}{2^{|L|-1}}$ & 0.4993 & 0.4876 & 0.4999\tabularnewline
$\frac{1}{\pi_{i}}$ & \textbf{0.5017} & 0.4989 & \textbf{0.5013}\tabularnewline
$\frac{1}{\log(1+\pi_{i})}$ & 0.4977 & 0.4998 & 0.4979\tabularnewline
\end{tabular}
\par\end{centering}

\caption{ERR scores w.r.t. weighting schemes in $\ell_{3}$ (Eq.\ref{eq:pointwise}),
and $\pi_{i}$ is the position of the object $o_{i}$ in the ranked
list. \label{tab:Effect-of-weighting-position}}
\end{table}

\textbf{Weighting schemes.} The importance of appropriate weighting
scheme for position-wise decomposition of loss function is reported
in Table~\ref{tab:Effect-of-weighting-position}. For loss functions
based on the $\wPL$ and the $\wPLL$, weighting is critical to achieve
high quality. For the $\rPL$, however, the role of weighting generally
does not help to improve performance. This is interesting since it
is probable that the reverse Plackett-Luce ($\rPL$) puts more emphasis
of removing the bad objects (see Eq.(\ref{eq:PL-reverse})), which
in effect, increases the chance of the good objects at the top.

Table~\ref{tab:Effect-of-weighting-pair} reports the performance
of models with weighted pairwise decomposition. Again, it clearly
demonstrates that with careful weighting schemes we can greatly improve
the quality of models. 

\begin{table}
\begin{centering}
\begin{tabular}{c|c|c|c|c}
Weight $V_{ij}^{(q)}$ & Quad. & Logist. & Hinge & $\wUB$\tabularnewline
\hline 
$1$ & \emph{0.4789} & \emph{0.4801} & \emph{0.4787} & \emph{0.4829}\tabularnewline
$\frac{1}{N_{q}}$ & 0.4881 & 0.4885 & 0.4864 & 0.4912\tabularnewline
$|r_{i}-r_{j}|$ & 0.4789 & 0.4834 & 0.4772 & 0.4854\tabularnewline
$\frac{|r_{i}-r_{j}|}{N_{q}}$ & 0.4883 & 0.4909 & 0.4885 & 0.4918\tabularnewline
$\frac{(R_{i}-R_{j})(\eta_{i}-\eta_{j})}{\mbox{NDC\ensuremath{G_{Max}}}}$ & 0.4928 & 0.4962 & 0.4935 & 0.4976\tabularnewline
$(R_{i}-R_{j})(\eta_{i}-\eta_{j})$ & 0.4916 & 0.4964 & 0.4917 & 0.4950\tabularnewline
$R_{i}-R_{j}$ & 0.4903 & 0.4936 & 0.4903 & 0.4939\tabularnewline
$\frac{R_{i}-R_{j}}{N_{q}}$ & \textbf{0.4952} & \textbf{0.4979} & \textbf{0.4953} & \textbf{0}.\textbf{4977}\tabularnewline
\end{tabular}
\par\end{centering}

\caption{ERR scores w.r.t. weighting schemes in ($\ell_{4}$, Eq.\ref{eq:pairwise}),
where $R_{i}=\frac{2^{r_{i}}-1}{2^{|L|-1}}$, $\eta_{i}=\frac{1}{\log(1+\pi_{i})}$,
and $\pi_{i}$ is the position of the object $o_{i}$ in the ranked
list. Note that in the pairwise approximation to MRF, the weight should
not include the ratings since they have been accounted for in the
log-likelihood. \label{tab:Effect-of-weighting-pair}}

\end{table}

\section{Discussion}

In this paper, we aim at taking a holistic view when designing a robust
learning to rank (LTR) algorithm. We consider domains of Social Question
Answering (SQA) and Web Information Retrieval (WIR). This section
presents some more elaboration on a variety of design aspects.

In SQA, we have chosen the separate representation of features, and
investigated two nonlinear rank functionals $f_{1}$ and $f_{2}$
against three losses: the multiclass logistic $\ell_{1}$, the loss
$\ell_{2}$ based on approximating the MRR metric, and the pairwise
logistic loss $\ell_{3}$. The results indicate that it is better
to use the loss specifically designed to the task. More specifically,
when the task is to predict the single best object, the multiclass
logistic loss should work better than more generic rank losses. We
note in passing that we do not aim at outperforming state-of-the-art
results in the QA literature, which has a history of several decades,
but rather show that simple representation such as words can be useful
for this complicated task.

In the WIR task, on the other hand, combined features are pre-computed,
and are not revealed to the public. However, it is plausible that
they contain relevance assessment accordingly multiple criteria. Although
the feature details are hidden, the quadratic rank functionals can
be useful to detect which feature conjunctions are predictive. Our
evaluation confirms that second-order features are indeed useful,
provided that we have effective method to control overfitting. In
particular, we employ regularisation and feature filtering for the
task. This result, together with those reported for the Yahoo! LTR
challenge \cite{chapelle2011yahoo}, suggests that the space of rank
functionals is a critical factor in achieving high performance LTR
algorithms.

Different from SQA, the WIR data contain relevance ratings in a small
numerical scale, which suggests the presence of ties. This calls for
group-level rank functionals, where objects of the same rating are
grouped into a mega-object. We have evaluated several aggregation
methods for computing the group-level rank functionals. The results
indicate that max and geometric mean aggregations can be competitive
in term of the ERR score. The main benefit is that learning can operate
directly on the level of groups, and since the number of groups is
often much smaller than the number of objects per query, computational
saving can be attained.

Like in the SQA task, we also evaluated the direct optimisation of
approximation of rank metrics, which has been suggested by several
recent studies \cite{qin2010general} \cite{valizadegan2010learning}
\cite{wu2009smoothing}. However, our empirical results show that
this is not necessarily the case. We conjecture that due to the approximation
of the step function, the resulting functions are often highly complex,
and possibly non-convex, making it difficult for optimisation. This
observation is shared in \cite{wu2009smoothing}.

The strong message we obtained from the experiments with the WIR data
is that for many piecewise loss functions ($\ell_{3}$in Eq.\ref{eq:pointwise}
and $\ell_{4}$ in Eq.\ref{eq:pairwise}), weighting is an important
factor. This is expected since piecewise loss functions often operate
locally with one or two free variables, while rank metrics are often
a function of the whole query. In particular, for the element-wise
decomposition of $\ell_{3}$, rank discount weighting is the most
influential (Table~\ref{tab:Effect-of-weighting-position}). For
the pairwise decomposition of $\ell_{4}$, rating difference combined
with query length normalisation is the most effective method (Table~\ref{tab:Effect-of-weighting-pair}).

\section{Related Work}

Several rank functionals other than the simple linear ones have been
suggested in the literature, notably the neural net \cite{burges2005learning},
kernels \cite{joachims2002optimizing}\cite{moschitti2010linguistic},
regression trees \cite{li2007mcrank} and bilinear \cite{bai2010learning}.
The neural net rank functionals are often non-convex, and their discriminative
power has not been clearly documented. The kernels, on the other hand,
can be expensive for large-scale data, since most kernel-based algorithms
scale super-linearly in number of training documents. Regression trees
are interesting due to their flexibility in function approximation.
Finally, bilinear functions are the linear combination of the query
and object feature vectors - thus it falls into the group of separate
feature representation. Our work contributes to this line of representation
by introducing several non-linear functions.

Instance weighting has been used in LTR models in several places \cite{burges2005learning}
\cite{cossock2008statistical} \cite{chen22ranking}. In \cite{burges2005learning},
the piecewise approximation in Eq.(\ref{eq:pairwise}) is used where
the function $\varphi(.)$ is the log loss of the logistic model,
and the weight $W_{ij}=\exp\{r_{i}\}/(\exp\{r_{i}\}+\exp\{r_{j}\})$.
In \cite{cossock2008statistical}, element-wise decomposition in Eq.(\ref{eq:pointwise})
is suggested to obtain a bound on the NDCG score, but the details
of $W_{i}$ are not reported. In \cite{chen22ranking}, NDCG-based
weights are introduced for pairwise and Plackett-Luce models. We extend
this by investigating a wider range of weighting schemes on several
new models.

Query-level models have been advocated in several places \cite{xia2008listwise}
\cite{liu2009learning} \cite{volkovs2009boltzrank}. In particular,
in \cite{xia2008listwise}, the standard Plackett-Luce model is employed,
but weighting is not considered. In \cite{volkovs2009boltzrank},
a Markov random field is suggested, and the learning involves MCMC
sampling, which leads to non-smooth risk functionals and is hard to
judge the convergence. Our weighted pseudo-likelihood and weighted
piecewise approximation are, on the other hand, smooth.

There have been a number of recent studies on how to optimise rank
metrics directly \cite{taylor2008softrank}\cite{xu2008directly}\cite{wu2009smoothing}
\cite{chakrabarti2008structured}\cite{chen22ranking}. There are
two approaches: one is based on approximating the rank metrics \cite{qin2010general}
\cite{valizadegan2010learning} \cite{wu2009smoothing}, and the other
on bounding the rank errors \cite{cossock2008statistical}\cite{chen22ranking}.
We contribute further to the first approach by approximating the ERR
and MRR metrics.

The issue of ranking with ties in the context of LTR has received
little attention \cite{zhou2008learning}\cite{Truyen:2011a}. In
\cite{zhou2008learning}, ties are considered among pairs of objects,
but not the entire group of objects. On the other hand, \cite{Truyen:2011a}
studies ties in groups, but their algorithm is only efficient for
a particular case of group-level rank functionals (geometric mean
- see Table~\ref{tab:Group-level-rank-functions.}).

\section{Conclusion}

In this paper, we have investigated choices when designing learning
to rank algorithm to perform well against evaluation criteria. We
evaluated design aspects on two tasks: answer ranking in a Social
Question Answering site, and Web Information Retrieval. Among others,
we have found that representing and selecting features, choosing a
(cost-sensitive) loss function, handling ties, and weighting data
instances are important to achieve high performance algorithms. 

LTR is a fast growing field with many established techniques, and
thus it is of practical importance to have a clear picture where nuts
and bolts are identified. This paper is aimed as a step towards that
goal. However, there are theoretical issues that needed to be addressed.
First, this is mathematically a function estimation problem, where
we still do not have a good understanding of the generalization properties.
Second, it appears that the structure of the rank functional space
is a key to the success of rank algorithms, and thus there is room
for more investigation into data partitioning and nonparametric settings.

\appendix

\section{Proof of The Upper-bound in $\wUB$}

Recall from Eq.(\ref{eq:MRF}) that $Z=\sum_{\r}\prod_{i}\prod_{j>i}\psi(r_{i},r_{j})$.
Let us define an extended function $\Psi_{ij}(\r)=\psi(r_{i},r_{j})$
for all realisations of $\r_{\neg ij}$, thus
\[
Z=\sum_{\r}\prod_{ij|j>i}\Psi_{ij}(\r)
\]
According to the general H\"{o}lder's inequality 
\begin{equation}
Z\le\prod_{ij|j>i}\left(\sum_{\r}\Psi_{ij}(\r)^{q_{ij}}\right)^{\frac{1}{q_{ij}}}\label{eq:Holder}
\end{equation}
for any $q_{ij}>0$ subject to $\sum_{ij|j>i}1/q_{ij}=1$. Further,
notice that $\Psi_{ij}(\r)$ indeed depends only on $(r_{i},r_{j})$,
thus 
\begin{eqnarray}
\sum_{\r}\Psi_{ij}(\r)^{q_{ij}} & = & \sum_{\r_{\neg ij}}\sum_{r_{i}}\sum_{r_{j}}\Psi_{ij}(\r)^{q_{ij}}\nonumber \\
 & = & |L|^{N-2}\sum_{r_{i}}\sum_{r_{j}}\psi(r_{i},r_{j})^{q_{ij}}\label{eq:appendix1}
\end{eqnarray}
where $|L|$ is the size of the label set. The factor $|L|^{N-2}$
comes from the fact that there are $|L|^{N-2}$ ways of enumerating
a set of $N-2$ discrete variables in $\r_{\neg ij}$, each of size
$|L|$.

For any integer $q_{ij}$, we have 
\begin{equation}
\sum_{r_{i}}\sum_{r_{j}}\psi(r_{i},r_{j})^{q_{ij}}\le\left(\sum_{r_{i}}\sum_{r_{j}}\psi(r_{i},r_{j})\right)^{^{q_{ij}}}\label{eq:poly-expansion}
\end{equation}
since $\psi>0$ and the LHS is a part in the expansion of the RHS.

Substituting (\ref{eq:poly-expansion}) into (\ref{eq:appendix1})
and then into (\ref{eq:Holder}), we obtain
\begin{eqnarray*}
Z & \le & \prod_{ij|j>i}\left(|L|^{\frac{N-2}{q_{ij}}}Z_{ij}\right)\,\,\mbox{or equivalently},\\
\log Z & \le & \sum_{ij|j>i}\log Z_{ij}+C
\end{eqnarray*}
where $Z_{ij}=\sum_{r_{i}}\sum_{r_{j}}\psi(r_{i},r_{j})$, and
\begin{eqnarray*}
C & = & \sum_{ij|j>i}\log|L|^{\frac{N-2}{q_{ij}}}=(N-2)\log|L|
\end{eqnarray*}
In the last equation, we have eliminated $q_{ij}$ by using the fact
that $\sum_{ij|j>i}1/q_{ij}=1$. 

Recall that $P(\r)=\prod_{ij|j>i}\psi(r_{i},r_{j})/Z$, thus
\begin{eqnarray*}
\ell(\r) & = & -\log P(\r)\\
 & = & \log Z-\sum_{ij|j>i}\log\psi(r_{i},r_{j})\\
 & \le & \sum_{ij|j>i}\left(\log Z_{ij}-\log\psi(r_{i},r_{j})\right)+C\\
 & =- & \sum_{ij|j>i}\log Q(r_{i},r_{j})+C
\end{eqnarray*}
where we have used $Q(r_{i},r_{j})=\psi(r_{i},r_{j})/Z_{ij}$. This
completes the proof.


\begin{thebibliography}{10}

\bibitem{bai2010learning}
B.~Bai, J.~Weston, D.~Grangier, R.~Collobert, K.~Sadamasa, Y.~Qi, O.~Chapelle,
  and K.~Weinberger.
\newblock Learning to rank with (a lot of) word features.
\newblock {\em Information retrieval}, 13(3):291--314, 2010.

\bibitem{burges2005learning}
C.~Burges, T.~Shaked, E.~Renshaw, A.~Lazier, M.~Deeds, N.~Hamilton, and
  G.~Hullender.
\newblock {Learning to rank using gradient descent}.
\newblock In {\em Proc. of ICML}, page~96, 2005.

\bibitem{chakrabarti2008structured}
S.~Chakrabarti, R.~Khanna, U.~Sawant, and C.~Bhattacharyya.
\newblock {Structured learning for non-smooth ranking losses}.
\newblock In {\em Proceeding of the 14th ACM SIGKDD international conference on
  Knowledge discovery and data mining}, pages 88--96. ACM, 2008.

\bibitem{chapelle2011yahoo}
O.~Chapelle and Y.~Chang.
\newblock Yahoo! learning to rank challenge overview.
\newblock In {\em JMLR Workshop and Conference Proceedings}, volume~14, pages
  1--24, 2011.

\bibitem{chapelle2009expected}
O.~Chapelle, D.~Metlzer, Y.~Zhang, and P.~Grinspan.
\newblock {Expected reciprocal rank for graded relevance}.
\newblock In {\em CIKM}, pages 621--630. ACM, 2009.

\bibitem{chen22ranking}
W.~Chen, T.Y. Liu, Y.~Lan, Z.~Ma, and H.~Li.
\newblock Ranking measures and loss functions in learning to rank.
\newblock {\em Advances in Neural Information Processing Systems}, 22:315--323.

\bibitem{cossock2008statistical}
D.~Cossock and T.~Zhang.
\newblock {Statistical analysis of Bayes optimal subset ranking}.
\newblock {\em IEEE Transactions on Information Theory}, 54(11):5140--5154,
  2008.

\bibitem{freund2004eba}
Y.~Freund, R.~Iyer, R.E. Schapire, and Y.~Singer.
\newblock An efficient boosting algorithm for combining preferences.
\newblock {\em Journal of Machine Learning Research}, 4(6):933--969, 2004.

\bibitem{jarvelin2002cumulated}
K.~J{\"a}rvelin and J.~Kek{\"a}l{\"a}inen.
\newblock {Cumulated gain-based evaluation of IR techniques}.
\newblock {\em ACM Transactions on Information Systems (TOIS)}, 20(4):446,
  2002.

\bibitem{joachims2002optimizing}
T.~Joachims.
\newblock {Optimizing search engines using clickthrough data}.
\newblock In {\em Proc. of SIGKDD}, pages 133--142. ACM New York, NY, USA,
  2002.

\bibitem{li2007mcrank}
P.~Li, C.~Burges, Q.~Wu, JC~Platt, D.~Koller, Y.~Singer, and S.~Roweis.
\newblock Mcrank: Learning to rank using multiple classification and gradient
  boosting.
\newblock {\em Advances in neural information processing systems}, 2007.

\bibitem{liu2009learning}
T.Y. Liu.
\newblock {Learning to rank for information retrieval}.
\newblock {\em Foundations and Trends in Information Retrieval}, 3(3):225--331,
  2009.

\bibitem{luce1959individual}
R.D. Luce.
\newblock {\em {Individual choice behavior}}.
\newblock Wiley New York, 1959.

\bibitem{moschitti2010linguistic}
A.~Moschitti and S.~Quarteroni.
\newblock {Linguistic kernels for answer re-ranking in question answering
  systems}.
\newblock {\em Information Processing \& Management}, 2010.

\bibitem{plackett1975analysis}
R.L. Plackett.
\newblock {The analysis of permutations}.
\newblock {\em Applied Statistics}, pages 193--202, 1975.

\bibitem{qin2010general}
T.~Qin, T.Y. Liu, and H.~Li.
\newblock A general approximation framework for direct optimization of
  information retrieval measures.
\newblock {\em Information retrieval}, 13(4):375--397, 2010.

\bibitem{taylor2008softrank}
M.~Taylor, J.~Guiver, S.~Robertson, and T.~Minka.
\newblock {SoftRank: optimizing non-smooth rank metrics}.
\newblock In {\em Proceedings of the international conference on Web search and
  web data mining}, pages 77--86. ACM, 2008.

\bibitem{Truyen:2011a}
T.~Truyen, D.Q Phung, and S.~Venkatesh.
\newblock Probabilistic models over ordered partitions with applications in
  document ranking and collaborative filtering.
\newblock In {\em Proc. of SIAM Conference on Data Mining (SDM)}, Mesa,
  Arizona, USA, 2011. SIAM.

\bibitem{valizadegan2010learning}
H.~Valizadegan, R.~Jin, R.~Zhang, and J.~Mao.
\newblock {Learning to Rank by Optimizing NDCG Measure}.
\newblock In {\em NIPS}, 2009.

\bibitem{volkovs2009boltzrank}
M.N. Volkovs and R.S. Zemel.
\newblock {BoltzRank: learning to maximize expected ranking gain}.
\newblock In {\em Proceedings of the 26th Annual International Conference on
  Machine Learning}. ACM New York, NY, USA, 2009.

\bibitem{wu2009smoothing}
M.~Wu, Y.~Chang, Z.~Zheng, and H.~Zha.
\newblock {Smoothing DCG for learning to rank: A novel approach using smoothed
  hinge functions}.
\newblock In {\em Proceeding of the 18th ACM conference on Information and
  knowledge management}, pages 1923--1926. ACM, 2009.

\bibitem{xia2008listwise_FULL}
F.~Xia, T.Y. Liu, J.~Wang, W.~Zhang, and H.~Li.
\newblock {Listwise approach to learning to rank: theory and algorithm}.
\newblock In {\em Proceedings of the 25th international conference on Machine
  learning}, pages 1192--1199. ACM, 2008.

\bibitem{xia2008listwise}
F.~Xia, T.Y. Liu, J.~Wang, W.~Zhang, and H.~Li.
\newblock {Listwise approach to learning to rank: theory and algorithm}.
\newblock In {\em Proc. of ICML}, pages 1192--1199, 2008.

\bibitem{xu2008directly}
J.~Xu, T.Y. Liu, M.~Lu, H.~Li, and W.Y. Ma.
\newblock {Directly optimizing evaluation measures in learning to rank}.
\newblock In {\em Proceedings of the 31st annual international ACM SIGIR
  conference on Research and development in information retrieval}, pages
  107--114. ACM, 2008.

\bibitem{zhou2008learning}
K.~Zhou, G.R. Xue, H.~Zha, and Y.~Yu.
\newblock {Learning to rank with ties}.
\newblock In {\em Proc. of SIGIR}, pages 275--282, 2008.

\end{thebibliography}
\end{document}